%% file: PRB/main.tex
\documentclass[%
 reprint,
 superscriptaddress,
 amsmath,amssymb,
 aps,
 longbibliography,
 prb,
]{revtex4-2}

\usepackage{graphicx}
\usepackage{dcolumn}
\usepackage{bm}
\usepackage{physics}
\usepackage{float}
\usepackage{xcolor}
\usepackage{soul}
\usepackage[normalem]{ulem}
\usepackage[colorlinks=true, linkcolor=black, citecolor=black, urlcolor=black]{hyperref}

\colorlet{blue}{blue!70!black} 
\colorlet{red}{red!70!black} 

\newcommand{\bb}[1]{{\color{black}{#1}}}

\newif\ifnature
\naturetrue

\def\oden{Oden Institute for Computational Engineering and Sciences, The University of Texas at Austin, Austin, TX 78712, USA}
\def\utphysics{Department of Physics, The University of Texas at Austin, Austin, TX 78712, USA}

\begin{document}

\preprint{APS/123-QED}

\title{\bb{Electron correlation in semiconductors and insulators via symbolic regression}}

\author{Nick Pant}
\email{npant@utexas.edu}
\affiliation{\oden}%
\affiliation{\utphysics}%

\author{Viet-Anh Ha}
\affiliation{\oden}%
\affiliation{\utphysics}%

\author{Donghwan Kim}
\affiliation{\oden}%
\affiliation{\utphysics}%

\author{Feliciano Giustino}
\email{fgiustino@oden.utexas.edu}
\affiliation{\oden}%
\affiliation{\utphysics}%

\date{\today}

\begin{abstract}
    Predicting quasiparticle energies in materials requires expensive numerical evaluations of the electron self-energy. This limits calculations to ordered systems with small unit cells. Here, using symbolic regression, we show that the GW self-energy \bb{can be accurately approximated with compact analytical functions of physically motivated Kohn-Sham descriptors}. These expressions can be learned from a single GW calculation in the ordered phase and remain accurate under symmetry breaking induced by quantum and thermal fluctuations, elastic deformations, and amorphous disorder. This development enables routine GW calculations of complex materials with thousands of atoms at a computational cost comparable to semi-local density functional theory. We demonstrate the accuracy of this approach for covalent semiconductors, ionic insulators, and two-dimensional materials. These results establish symbolic regression as a viable route to predictive, interpretable, and transferable many-body electronic structure models.
\end{abstract}
\maketitle

\section{Introduction}

The GW approximation of many-body perturbation theory provides an accurate \textit{ab initio} description of electron correlation in materials. This is critical for predicting quasiparticle properties, including band gaps, effective masses, and all other properties derived from the band structure~\cite{hybertsen1986electron, onida2002electronic, golze2019gw}. However, its practical application is limited by the high computational cost associated with evaluating the dynamical, non-local self-energy operator.

Recent work has shown that machine learning (ML) can accelerate GW calculations~\cite{Westermayr2021Physically, knosgaard2022representing, zauchner2023accelerating, hou2024unsupervised, dong2024equivariant, venturella2025unified}, often by learning corrections to less expensive density functional theory (DFT) calculations. However, two key challenges must be addressed before ML-based approaches can be widely deployed for many-body calculations with accuracy comparable to state-of-the-art electronic structure methods. First, the training requirements are daunting~\cite{xu2023small}, typically demanding thousands of examples to achieve acceptable accuracy. This is problematic because many-body calculations are extremely costly and require large-scale supercomputing resources even for simple systems. Second, existing models suffer from a lack of interpretability~\cite{zhong2022explainable}. Black-box architectures such as neural networks or models based on complex engineered features obscure how predictions are actually made. This makes it difficult to explain predictions and assess their reliability. Without a clear understanding of what the model learns, it is hard to predict how it will generalize outside the training set.

An alternative to this black-box approach is to use artificial intelligence (AI) to learn governing equations from the data. This approach, known as symbolic regression~\cite{makke2024interpretable}, has been applied in diverse fields from dynamical systems to materials informatics, and successfully rediscovered known laws and revealed new analytical models~\cite{schmidt2009distilling, brunton2016discovering, cranmer2020discovering, ouyang2018sisso, wang2019symbolic, Ma2022Evolving}. Unlike black-box models, symbolic regression yields explicit analytical expressions that are interpretable and can often be trained effectively even from limited datasets.

In this work, we introduce a framework, which we refer to as symbolic GW (symGW), that distills the essential physics of GW quasiparticle corrections into compact equations depending only on DFT quantities. We show that symGW can predict, at the cost of semi-local DFT calculations, GW quasiparticle energies in regimes that are exceptionally demanding for traditional methods: GW band structures at finite temperature, strain-dependent GW bands of two-dimensional (2D) atomic crystals, and GW electronic spectra of amorphous solids. We also show a simple symGW equation whose functional form is shared by materials of diverse chemistries, and which has an intuitive physical interpretation.

The manuscript is organized as follows. In Sec.~\ref{sec:symGW}, we describe how to perform symbolic regression of quasiparticle corrections using physically informed features derived from DFT calculations. In Sec.~\ref{sec:results}, we present the main results demonstrating the predictive power of symGW. Sections~\ref{sec:finite-temp}, \ref{sec:strain}, and \ref{sec:amorphous} demonstrate the application of symGW to predicting band renormalization in diamond due to quantum and thermal fluctuations, strain-induced modifications of the intervalley energy and conductivity effective mass in monolayer MoS$_2$, and quasiparticle corrections in amorphous SiO$_2$, respectively. Finally, in Sec.~\ref{sec:discussion}, we perform feature analysis to examine why symGW generalizes well to unseen structures. We also discuss additional physical constraints to address the non-uniqueness of the symbolic expressions and demonstrate a unified form that describes quasiparticle corrections across diverse solids. Technical details of our calculations are provided in the Appendix.

\section{Symbolic regression of quasiparticle corrections}\label{sec:symGW}

The self-energy $\Sigma$ encodes the effects of electronic interactions beyond mean-field theory~\cite{hedin1965new}. Quasiparticle energies are obtained by replacing the DFT exchange--correlation (XC) potential with the self-energy, $E_{n\mathbf{k}} = \varepsilon_{n\mathbf{k}} + \bra{\psi_{n\mathbf{k}}}\,(\Sigma - V^\text{XC})\,\ket{\psi_{n\mathbf{k}}}$, where $n$ is the band index, $\mathbf{k}$ is the crystal momentum, $E_{n\mathbf{k}}$ is the quasiparticle energy, $\varepsilon_{n\mathbf{k}}$ is the Kohn--Sham (KS) eigenvalue, and $V^\text{XC}$ is the XC potential~\cite{hybertsen1986electron}. Within the GW method, the self-energy is approximated as $\Sigma = iGW$, where $G$ is the electron Green's function and $W$ is the screened Coulomb interaction; this approximation systematically corrects the band-gap underestimation in DFT, and predicts band structures in excellent agreement with experiments for weakly- to moderately-correlated materials~\cite{hybertsen1986electron, van2006quasiparticle, shishkin2007accurate}.  

Since GW quasiparticle corrections are obtained by taking the expectation value of $\Sigma - V^\text{XC}$ on KS states, we represent KS states using a minimal set of identifiers that are easily obtained from DFT calculations, and we seek to discover equations that yield GW corrections as explicit functions of such identifiers. This approach builds on the observation that, in certain cases, quasiparticle corrections can be expressed as polynomials of KS eigenvalues~\cite{yazyev2012quasiparticle, waroquiers2013band}.

The simplest identifiers that can be obtained from a DFT calculation are the expectation values of the kinetic energy ($t_{n\mathbf{k}}$), the external potential from the ionic lattice ($v^{\text{ext}}_{n\mathbf{k}}$), the Hartree potential ($v^{\text{H}}_{n\mathbf{k}}$), and the XC potential ($v^{\text{XC}}_{n\mathbf{k}}$); these identifiers add up to the KS eigenvalue, $\varepsilon_{n\mathbf{k}} = t_{n\mathbf{k}} + v^{\text{ext}}_{n\mathbf{k}} + v^{\text{H}}_{n\mathbf{k}} + v^{\text{XC}}_{n\mathbf{k}}$. Using these identifiers, we define the Kohn-Sham-4 (KS4) vector,  $\mathbf{x}_{n\mathbf{k}} = [\,t_{n\mathbf{k}},\, v^{\text{H}}_{n\mathbf{k}},\, v^{\text{XC}}_{n\mathbf{k}},\, \varepsilon_{n\mathbf{k}}\,]$, which we adopt as the feature set for symbolic regression. With this choice, all discovered expressions will be a function of at least one of these four identifiers. By convention, we refer eigenvalues in the KS4 vector to the valence band maximum; the average Hartree potential in the unit cell is set to zero, following standard practice in electronic structure codes. We note that the KS4 vector can readily be obtained by postprocessing the outputs of a standard DFT calculation. Other more advanced features could be used for symbolic regression, but in the following we show that this minimalist choice already carries strong predictive power.

\begin{figure}[htp!]
    \centering
    \includegraphics{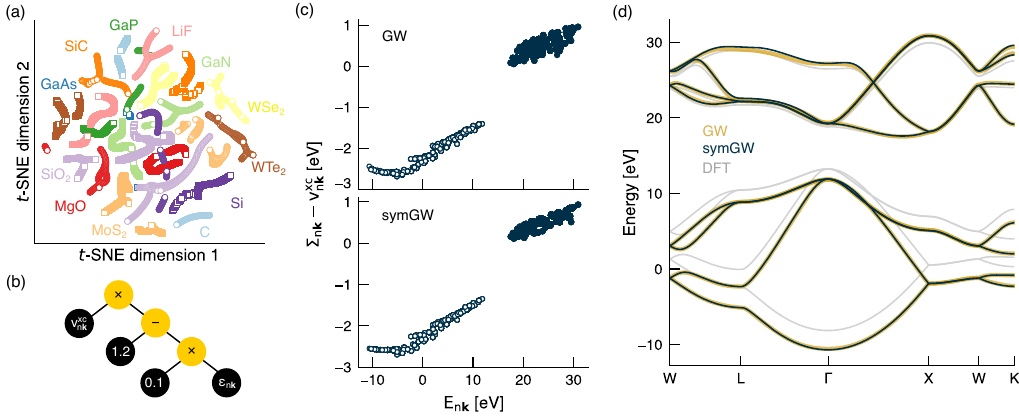}
    \caption{
            (a)~Two-dimensional $t$-SNE embedding of the KS4 vectors of the highest valence band and lowest conduction band for 12 materials as indicated by labels. Circles denote conduction states and squares denote valence states.
            (b)~Example binary tree representing a candidate expression, $\Sigma_{n\textbf{k}} = v^\text{XC}_{n\textbf{k}} (1.2 - 0.1 \varepsilon_{n\textbf{k}})$, in the evolutionary search of symGW. 
            (c)~Comparison of explicit GW quasiparticle corrections computed from first principles with the corrections obtained from symGW [Eq.~\eqref{eq:diamond}] for diamond. Filled disks denote conduction states,  circles denote valence states. The parameters used in Eq.~\eqref{eq:diamond} are $\alpha_1= 0.187$, $\alpha_2=1.338$, $\alpha_3=0.470$, $\alpha_4=0.872$, $\alpha_5=0.0306$, and $\alpha_6=0.0653$. 
            (d)~Comparison of Wannier-interpolated GW band structure (yellow), symGW band structure (dark blue), and DFT band structure (gray) for diamond. The model is trained on $G_0W_0$ corrections computed on a uniform Brillouin zone grid with 8$\times$8$\times$8 points.}
    \label{fig:symGW}
\end{figure}

Figure~\ref{fig:symGW}(a) shows how the KS4 vector represents a highly selective chemical fingerprint for KS states. We applied $t$-distributed stochastic neighbor embedding (t-SNE) \cite{maaten2008visualizing} to the KS4 vectors of 12 materials, which include covalent and ionic solids, narrow and wide band-gap materials, as well as bulk and 2D crystals. This low-dimensional representation reveals grouping of KS states by material type; furthermore, for each material, conduction and valence band manifolds are also distinctly separated, even though this information was not supplied as labels. These distinct regions indicate that the KS4 vector embeds detailed chemical information, and therefore represents a meaningful fingerprint of KS states. 

Importantly, since KS eigenvalues are continuous and differentiable functions of the ionic coordinates and the crystal momentum (for non-degenerate states), so is the KS4 vector; this vector is also invariant with respect to both crystal symmetries and lattice translations. In the presence of degeneracies, continuity, differentiability, and invariance hold for the sum of KS4 vectors over the degenerate manifold, owing to the invariance of the trace under unitary transformations. These properties make the KS4 vector a robust and physically grounded representation for learning the mathematical patterns underlying quasiparticle corrections.

To discover equations, we employ a global evolutionary search performed on binary expression trees, whereby candidate expressions evolve through mutation and crossover using \texttt{PySR}~\cite{cranmer2023interpretable}. Figure~\ref{fig:symGW}(b) illustrates the binary-tree representation of a candidate expression. Mutations introduce random changes to nodes in the tree while crossovers recombine sub-trees from two parent expressions to generate new candidates. The discovery of symGW equations is performed as a post-processing step following a standard GW calculation. Jupyter notebooks demonstrating this process are available on \texttt{Materials Cloud}~\cite{pant2026MatCloud} and \texttt{Code Ocean}~\cite{pant2026cloudocean}. After a few hundred evolutionary iterations, the search yields a set of the best-performing expressions, producing a Pareto front that balances accuracy and complexity. For completeness, in Tabs.~\ref{tab:diamond_conduction} and~\ref{tab:diamond_valence}, we report the Pareto front of expressions obtained for diamond. In our tests, we observed that simple expressions, albeit slightly less accurate, are easier to interpret and generalize better; conversely, complex expressions are more accurate for the training set, but are harder to explain and do not generalize well. Based on these observations, we select the simplest expression that achieves a desired level of accuracy. For all materials tested, we have found that polynomials are sufficient to accurately describe the self-energy.

Taking diamond as an illustrative example, we perform a GW calculation in the primitive cell, and from this calculation we learn the following equation:
\begin{eqnarray}\label{eq:diamond}
    \Sigma_{n\textbf{k}} &= &\{(v_{n\textbf{k}}^\text{XC}) ^4 [\varepsilon_{n\textbf{k}} - (\varepsilon_{n\textbf{k}} + \alpha_1)^3 - \alpha_2] - \alpha_3\}\, \theta(-\varepsilon_{n\textbf{k}}) \nonumber\\
    &+& (\alpha_4 v_{n\textbf{k}}^\text{XC} + \alpha_5 \varepsilon_{n\textbf{k}} - \alpha_6)\, \theta(\varepsilon_{n\textbf{k}})~.
    \label{eq:diamond}
\end{eqnarray}
All quantities are in Hartree atomic units with the step function defined as $\theta(x)=1$ for $x\ge 0$, and $0$ otherwise. The parameters $\alpha_i$ are given in the caption of Fig.~\ref{fig:symGW}. Despite being trained on the entire set of KS4 vectors, this expression only contains the KS eigenvalue and the expectation value of the XC potential. Figure~\ref{fig:symGW}(c) shows that Eq.~\eqref{eq:diamond} captures both the qualitative and quantitative features of quasiparticle corrections of KS states across both conduction and valence manifolds. Figure~\ref{fig:symGW}(d) further shows that this equation accurately reproduces the full GW band structure, even though only data from a coarse uniform grid were included in the training. In Supplemental Fig. S1~\cite{SM_note}\nocite{PhysRevB.43.1993, monserrat2016correlation, antonius2014many, monserrat2014extracting, logothetidis1992origin, clark1964intrinsic, cardona2005electron, ponce2015temperature}, we show how this equation also generalizes to other pseudopotentials and XC functionals that were not included in the training.\\

\section{Transferability across materials structures}\label{sec:results}
\subsection{Quantum fluctuations and finite temperature effects}\label{sec:finite-temp}

\begin{figure}[htp!]
    \centering
    \includegraphics{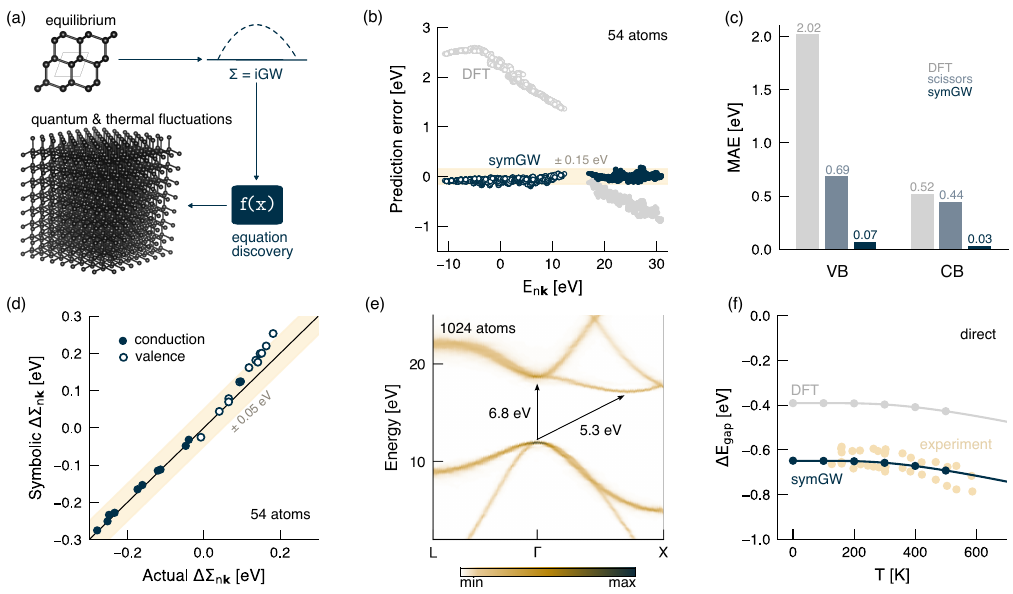}
    \caption{
            (a) The symGW equation learned from $G_0W_0$ corrections in the primitive unit cell of diamond can predict quasiparticle energies in supercells with quantum zero-point fluctuations and thermal disorder. 
            (b) Prediction error of DFT and symGW for quasiparticle energies in a 54-atom 3$\times$3$\times$3 supercell of diamond, including quantum zero-point fluctuations via the special displacement method~\cite{zacharias2020theory}. Filled disks denote conduction states and open circles denote valence states.
            (c) Mean absolute error comparison between DFT, DFT with a rigid scissors shift, and symGW.
            (d) Differences in self-energies between pristine and distorted 3$\times$3$\times$3 supercells (horizontal axis) compared to values predicted from the linear response of Eq.~\eqref{eq:diamond} (vertical axis), for states within 3~eV of the band edges.
            (e) symGW band structure of diamond including quantum zero-point renormalization. We use the special displacement method~\cite{zacharias2020theory} and a 1,024-atom 8$\times$8$\times$8 supercell containing 4,096 electrons. The color bar indicates the spectral weight. The calculated direct and indirect band gaps are 6.8~eV and 5.3~eV, respectively; these values are in close agreement with the experimental values of 7.0--7.1~eV~\cite{logothetidis1992origin} and 5.4~eV~\cite{clark1964intrinsic}, respectively.
            (f) Temperature dependence of the direct band gap of diamond, calculated using symGW and the special displacement method, compared to experiments~\cite{logothetidis1992origin}. The solid curves correspond to Bose-Einstein oscillator fits. We do not consider thermal lattice expansion, which is negligible for diamond in this temperature range~\cite{jacobson2019thermal}. Experimental data are referred to a nominal gap without electron-phonon effect of 7.73~eV, following previous work~\cite{antonius2014many, monserrat2016correlation}.}
    \label{fig:diamond}
\end{figure}

In Sec.~\ref{sec:symGW}, we demonstrated that an equation discovered for diamond on the coarse grid [Eq.~\eqref{eq:diamond}] can predict quasiparticle corrections at arbitrary \textbf{k} points in the Brillouin zone. Given that the GW self-energy in Eq.~\eqref{eq:diamond} is a smooth function of the KS4 vector, we hypothesize that it should be able to predict quasiparticle energies even after the symmetries of the original system are broken. To test this hypothesis, we investigate the quantum zero-point renormalization and the temperature dependence of the band structure of diamond using large supercells including quantum and thermal fluctuations, as illustrated in Fig.~\ref{fig:diamond}(a).

As a sanity check, we first perform explicit GW calculations on an affordable 54-atom supercell, with atoms displaced according to the thermal population of each phonon mode~\cite{zacharias2020theory}. Figure~\ref{fig:diamond}(b) shows that, while DFT eigenvalues deviate from GW quasiparticle energies by up to 2.5~eV, Eq.~\eqref{eq:diamond} predicts quasiparticle corrections within 0.1~eV of explicit GW calculations, across both conduction and valence manifolds. 
In Fig.~\ref{fig:diamond}(c), we quantify the average error across the Brillouin zone, showing that not only symGW outperforms DFT as expected, but it also improves significantly upon scissor-corrected DFT. Moreover, differentiating Eq.~\eqref{eq:diamond} with respect to ionic coordinates yields a symbolic model for the linear response of the self-energy~\cite{li2019electron}. As shown in Fig.~\ref{fig:diamond}(d), this linearized form closely predicts the actual variation of the self-energy due to ionic distortions.

Having passed these sanity checks, we now connect our calculations with experiments. To this end, we construct large diamond supercells with 1,024 atoms; this large size is needed when using the special displacement method~\cite{zacharias2020theory}. GW calculations at this large scale are very rarely performed due to their extreme computational cost~\cite{Govoni2015LargeGW, Vlcek2018swift, Wilhelm2018GW, Duchemin2021GW, ben2020accelerating, Gao2024Efficient}. 
Figure~\ref{fig:diamond}(e) shows the symGW band structure of diamond including quantum zero-point renormalization, yielding direct and indirect gaps in close agreement with experiment~\cite{clark1964intrinsic, logothetidis1992origin}. When we track the direct gap as a function of temperature [Fig.~\ref{fig:diamond}(f)], we find that symGW closely reproduces experimental measurements~\cite{monserrat2014extracting}, whereas DFT underestimates the phonon-induced renormalization of the direct band gap. These findings are in excellent agreement with explicit GW calculations (Supplemental Fig. S2). Similar considerations apply to the indirect gap, which is shown in Supplemental Fig. S3. The failure of DFT originates from the overscreening of the electron–phonon interaction~\cite{giustino2017rmp}; this error is corrected by the GW approximation~\cite{antonius2014many, monserrat2016correlation, faber2015exploring, li2019electron}, and the present symGW approach accurately captures the correction at the computational cost of a standard DFT calculation.

\subsection{Strain effects}\label{sec:strain}

As a second illustration of the ability of symGW to generalize beyond the original training set, we consider the electronic structure of strained MoS$_2$, a prototypical 2D semiconductor that is central to next-generation electronics~\cite{liu2021promises}, valleytronics~\cite{schaibley2016valleytronics}, and optoelectronics~\cite{Mak2016}. Strain is a powerful knob to tune the electronic properties of MoS$_2$, but it breaks crystalline symmetries, making explicit GW calculations very expensive. 

\begin{figure}[htp!]
    \centering
    \includegraphics{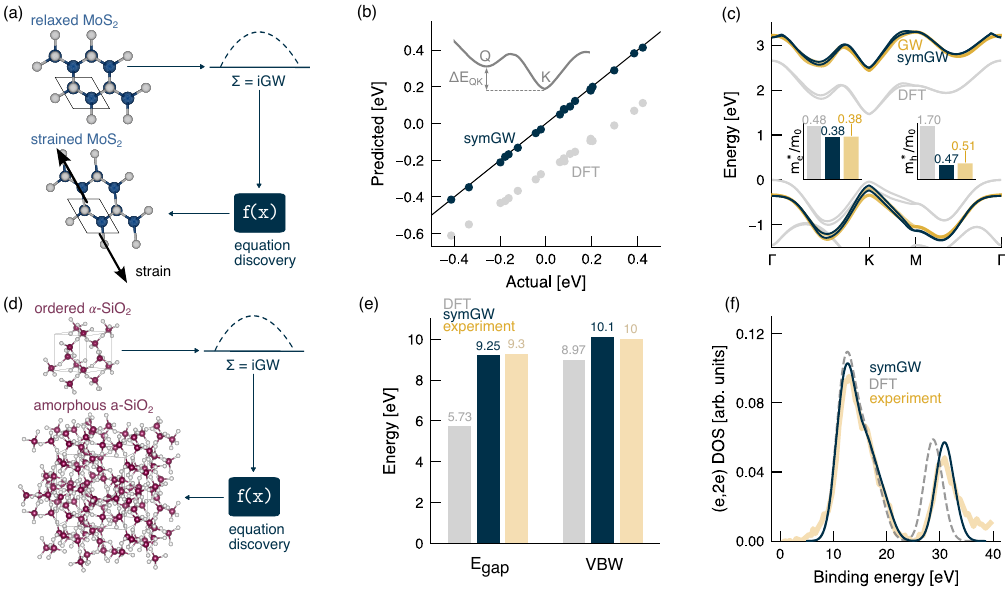}
    \caption{
            (a) The symGW equation learned from the high-symmetry unstrained MoS$_2$ crystal can predict quasiparticle energies in strained low-symmetry structures. The parameters appearing in Eq.~\eqref{eq:MoS2} are $\alpha_1=0.0451$, $\alpha_2=0.0774$, and $\alpha_3=1.0287$.
            (b) Comparison of DFT and symGW predictions of the Q--K conduction valley separation in MoS$_2$, compared to explicit GW calculations for 24 strained configurations. The diagonal line marks where predicted and actual values coincide.
            (c) Band structure of a representative strained configuration of MoS$_2$, including spin-orbit coupling, using DFT, symGW, and GW; the inset shows conductivity effective masses~\cite{hautier2014does} of holes and electrons at different levels of theory. 
            (d) The symGW equation trained on quartz ($\alpha$-SiO$_2$) can predict quasiparticle energies in the amorphous phase (a-SiO$_2$).
            (e) Comparison between DFT and symGW calculations of the band gap and the valence bandwidth (VBW) of a-SiO$_2$, compared to experiment~\cite{weinberg1979transmission}.
            The parameters appearing in Eq.~\eqref{eq:SiO2} are $\alpha_1=0.128$, $\alpha_2=0.111$, and $\alpha_3=0.254$.
            (f) Valence density of states of a-SiO$_2$ from symGW and DFT, compared with (e,2e) spectroscopy measurements~\cite{Fang1998Valence}. The theoretical curves are shifted to align with the leftmost peak. Higher binding energies correspond to deeper valence states. In Supplemental Fig. S4, we quantify the error of symGW compared to explicit GW calculations.}
    \label{fig:app}
\end{figure}

We ask whether an equation discovered for the high-symmetry undistorted structure could predict the electronic structure of this system under strain, see Fig. \ref{fig:app}(a). For unstrained MoS$_2$, we find the following equation: 
\begin{eqnarray}
    \Sigma_{n\textbf{k}} &= &(v_{n\textbf{k}}^\text{XC} + \alpha_1 t_{n\textbf{k}} - \alpha_2) \theta(-\varepsilon_{n\textbf{k}}) \nonumber\\
    & +& (\alpha_3 v_{n\textbf{k}}^\text{XC} + (\alpha_3-1) t_{n\textbf{k}}) \theta(\varepsilon_{n\textbf{k}}), \label{eq:MoS2}
\end{eqnarray}
with the parameters $\alpha_i$ provided in the caption of Fig.~\ref{fig:app}. Unlike Eq.~\eqref{eq:diamond}, in this case the kinetic energy component of the KS4 vector is essential to correctly capture GW corrections.

To investigate whether this equation generalizes under elastic deformation, we uniformly sample strains from $-3\%$ to $+3\%$, including both normal strain and shear strain components. A key performance metric in the description of MoS$_2$ under strain is the energy difference between the K and Q conduction valleys, which governs electron scattering rates~\cite{datye2022strain}. As shown in Fig.~\ref{fig:app}(b), DFT fails to predict this intervalley separation and often even reverses the valley ordering; in contrast, Eq.~\eqref{eq:MoS2} faithfully reproduces GW results. Figure ~\ref{fig:app}(c) further demonstrates that Eq.~\eqref{eq:MoS2} captures the full GW band structure for a representative strained configuration, and predicts the hole effective mass, a critical parameter for carrier mobility~\cite{ponce2020first}, within 8\% of the ground truth. On the other hand, not only DFT underestimates the band gap, but it also severely overestimates the hole effective mass by more than a factor of two.\\

\subsection{Amorphous disorder}\label{sec:amorphous}

Encouraged by the successes of Sec.~\ref{sec:finite-temp} and~\ref{sec:strain}, we push the present approach to its limits by testing whether it can generalize from a perfect crystal to a disordered amorphous solid. Amorphous materials pose a formidable challenge to GW calculations since the lack of long-range order requires supercells with hundreds to thousands of atoms~\cite{Liu2025Amorphous}. These materials are technologically crucial, as they are used in displays~\cite{Yu2016Metal}, solar cells~\cite{Liu2022Light}, and memory devices~\cite{Wang2025Amorphous}. Structural studies show that amorphization of high-symmetry materials leads to the breaking of long-range translational and rotational symmetries, while short-range order such as nearest-neighbor coordination is often preserved~\cite{Messmer1981Types}. Based on this observation, we hypothesize that symGW equations derived from the ordered crystalline phase may remain valid in the amorphous phase. This transferability would be consistent with the principle of nearsightedness of electronic matter~\cite{Prodan2005Nearsightedness}. 

To test this hypothesis, we consider amorphous SiO$_2$, the archetypal network-forming glass~\cite{Sarnthein1995Model, MartinSamos2010Unraveling}. We find the following equation for the quartz phase, $\alpha$-SiO$_2$:
\begin{eqnarray}
    \Sigma_{n\textbf{k}} &= &(v_{n\textbf{k}}^\text{XC} + \alpha_1 \varepsilon_{n\textbf{k}} - \alpha_2)\theta(-\varepsilon_{n\textbf{k}}) \nonumber\\
    & +& [v_{n\textbf{k}}^\text{XC}(1 - \alpha_3 \varepsilon_{n\textbf{k}})]\theta(\varepsilon_{n\textbf{k}}),
    \label{eq:SiO2}
\end{eqnarray}
with the parameters $\alpha_i$ provided in the caption of Fig.~\ref{fig:app}. Figure~\ref{fig:app}(d) shows that the symGW equation obtained from $\alpha$-SiO$_2$ transfers directly to the amorphous phase, a-SiO$_2$. While DFT underestimates both the band gap and the valence-band width in a-SiO$_2$, symGW predicts both quantities in excellent agreement with experiment~\cite{weinberg1979transmission, evrard1982photoelectric, keybus1986nonresonant, Fang1998Valence}, see Fig.~\ref{fig:app}(e). Beyond the band gap and band width, Fig.~\ref{fig:app}(f) shows that symGW accurately captures the complete electronic density of states: the valence-band DOS closely matches experimental measurements from secondary electron energy loss coincidence (e,2e) spectroscopy~\cite{Fang1998Valence}. 

\section{Discussion}\label{sec:discussion}

\begin{figure}[htp!]
    \centering
    \includegraphics{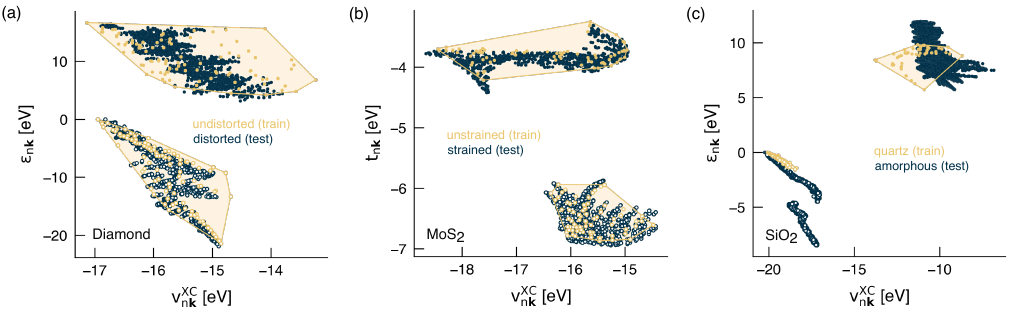}
    \caption{Distributions of the training and test data in feature space. (a) KS4 vectors from the undistorted diamond structure (yellow circles), projected onto the subspace of variables appearing in Eq.~\eqref{eq:diamond}, and KS4 vectors from the distorted diamond structure (blue circles), projected onto the same subspace. The shaded regions enclosed by the solid lines correspond to the convex hulls formed by the training points. Empty circles denote valence states, and filled circles denote conduction states. (b) Same as in panel (a), but for MoS$_2$ and Eq.~\eqref{eq:MoS2}. (c) Same as in panel (a), but for SiO$_2$ and Eq.~\eqref{eq:SiO2}.}
    \label{fig:ks4}
\end{figure}

\bb{The ability of symGW to learn equations across diverse materials and generalize to large structural deformations raises a natural question: why does it work? Fig.~\ref{fig:ks4}(a) shows the KS4 vectors for the diamond training data projected onto the subspace of Kohn-Sham descriptors that appear in Eq.~\eqref{eq:diamond}. It also displays the convex hull formed by the training points, together with the points from the distorted structure projected onto the same feature space. Although structural deformations shift the KS4 vectors in this space, these points largely remain within the convex hull of the training samples. This helps explain why learning the self-energy from a single undistorted structure is sufficient to predict quasiparticle energies in distorted structures. Similar considerations apply to MoS$_2$ and Eq.~\eqref{eq:MoS2}, as shown in Fig.~\ref{fig:ks4}(b). The case of a-SiO$_2$ is more surprising: Eq.~\eqref{eq:SiO2} remains highly predictive even though Fig.~\ref{fig:ks4}(c) shows that the KS4 vectors of a-SiO$_2$ have significantly smaller overlap with the convex hull formed by the quartz training points. This suggests that the analytic structure learned by symbolic regression supports not only interpolation, but also extrapolation under an appreciable distributional shift in the KS4-vector space.}

\bb{Despite the strong numerical performance of the learned expressions, a fundamental conceptual challenge remains: the expressions are not unique. This non-uniqueness reflects the fact that symbolic regression is an underdetermined problem. We propose that additional physical constraints may provide a systematic way to navigate this large space of possible expressions. One possible constraint is to ask whether a common functional form emerges across materials with diverse chemistries.}

By examining the candidate expressions for diamond, MoS$_2$, and SiO$_2$, we find that indeed the simplest expressions on the Pareto front share the same functional form. Despite their reduced complexity, these expressions still achieve a useful level of accuracy for practical calculations. Table~\ref{tab:rescaled_XC} shows that representing the self-energy as a rescaled XC energy,
$\Sigma_{n\mathbf{k}} = \alpha_1 v^\text{XC}_{n\mathbf{k}} + \alpha_2$,
already yields mean absolute errors of order 0.1~eV for a range of different materials including group IV and III--V semiconductors, covalent and ionic metal oxides, alkali halides, and 2D transition metal dichalcogenides. Including an additional linear dependence on the KS eigenvalue and kinetic energy,
\begin{equation} 
 \Sigma_{n\mathbf{k}} = \alpha_1 v^\text{XC}_{n\mathbf{k}} + \alpha_2 + \alpha_3 \varepsilon_{n\mathbf{k}} + \alpha_4 t_{n\mathbf{k}},
 \label{eq:shared}
\end{equation}
further reduces the MAE to the order of 10~meV for these diverse materials, as shown in Tab.~\ref{tab:linear_selfen}. 

\begin{figure}
    \centering
    \includegraphics{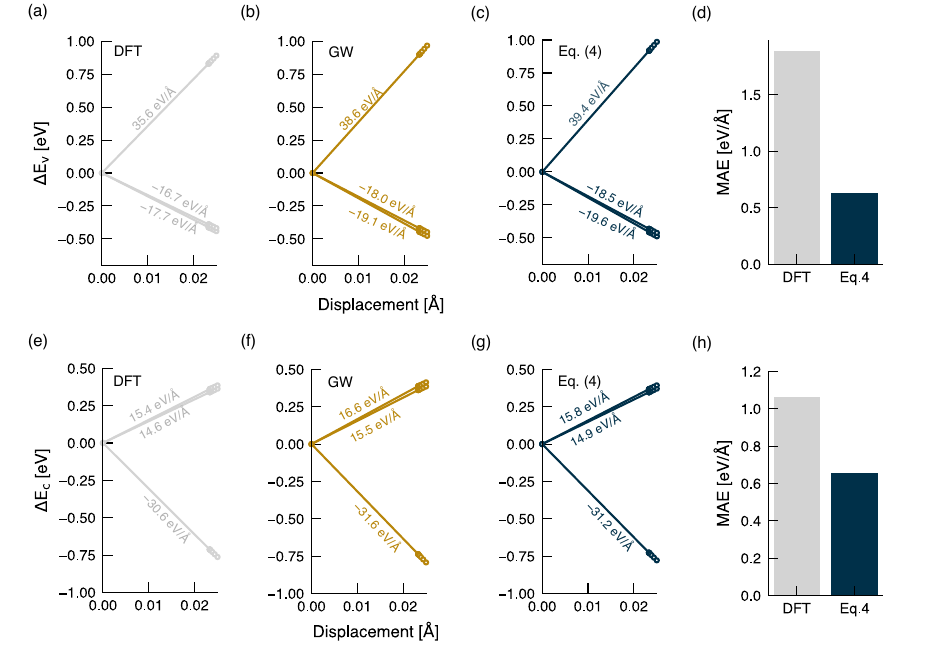}
    \caption{Linear response of the GW self-energy to atomic displacements for diamond. Splitting of the three-fold degenerate valence-band top at the $\Gamma$-point due to a frozen optical phonon with (a) DFT, (b) GW, and (c) Eq.~(4). (d) The mean absolute error for the slopes of panels a and c, with GW [panel (b)] as the ground truth.  (e)--(h) The same data as for panels (a)--(d) but for the splitting of the three-fold conduction-band bottom. The atomic displacements are obtained from $\Delta \tau_{\kappa\alpha} = e_{\kappa\alpha} u$, where $\kappa$ denotes the C atom in the unit cell, $\alpha$ is the Cartesian direction, $e_{\kappa\alpha}$ is the normalized, dimensionless vibrational eigenmode for the highest-frequency optical mode at $\Gamma$, and $u$ is the displacement amplitude.}
    \label{fig:derivatives}
\end{figure}

Perhaps more importantly, Eq.~\eqref{eq:shared} carries an intuitive meaning: the self-energy is described by a renormalization of the XC energy, together with dynamical corrections captured by $\varepsilon_{n\mathbf{k}}$ and $t_{n\mathbf{k}}$. Figure~\ref{fig:derivatives} shows that this unified form also robustly generalizes to linear-response calculations, opening the way to symbolic learning of many-body calculations of electron-phonon matrix elements. With access to larger GW datasets, it may soon become possible to investigate whether this equation applies to even broader classes of materials, and \bb{whether the associated coefficients can also be expressed in terms of learnable materials properties.}

Taken together, the present findings demonstrate that equation discovery via symbolic regression enables many-body GW calculations for complex materials that are beyond the reach of conventional methods, at the same cost of standard DFT calculations, and starting from very small datasets. Extending this framework to chemically heterogeneous systems would open many promising directions, including routine many-body calculations of alloys, defects, and heterostructures. Furthermore, the central idea underpinning the present approach may be generalizable to more complex many-body phenomena, spanning excitons, electron-phonon couplings, and strong correlations.

\vspace{10pt}
The authors thank Marios Zacharias for fruitful discussions. This work was supported by the Computational Materials Science program of the U.S. Department of Energy, Office of Science, Basic Energy Sciences, through award no. DE-SC0020129. Computational resources were provided by the National Energy Research Scientific Computing Center (a DOE Office of Science User Facility supported under Contract No.~DE-AC02-05CH11231) and the Texas Advanced Computing Center (TACC) at The University of Texas at Austin.

All data presented in this manuscript have been uploaded on the Materials Cloud database at this \href{https://archive.materialscloud.org/records/knmw1-sr498}{\ul{link}}. The codes used in this work, namely \texttt{EPW}~\cite{lee2023electron}, \texttt{Quantum ESPRESSO}~\cite{giannozzi2017advanced, giannozzi2020quantum}, \texttt{Wannier90}~\cite{Pizzi2020}, \texttt{BerkeleyGW}~\cite{deslippe2012berkeleygw}, and \texttt{PySR}~\cite{cranmer2023interpretable} are all open-source software and are freely available from their respective websites. Jupyter Notebooks for all calculations reported in this manuscript as well as code to print the KS4 vectors are publicly available on Code Ocean (\href{https://codeocean.com/capsule/8477105/tree/v1}{\ul{link}}), and are also provided through the Materials Cloud database (\href{https://archive.materialscloud.org/records/knmw1-sr498}{\ul{link}}).

\onecolumngrid

\clearpage

\clearpage

\clearpage

\clearpage
\twocolumngrid

\clearpage
\appendix
\input{PRB/appendix}

\bibliography{references}

\end{document}


\title{Supplemental Materials\\[5pt] Electron correlation in semiconductors and insulators via symbolic regression}

\author{Nick Pant}
\email{npant@utexas.edu}
\affiliation{\oden}%
\affiliation{\utphysics}

\author{Viet-Anh Ha}
\affiliation{\oden}%
\affiliation{\utphysics}%

\author{Donghwan Kim}
\affiliation{\oden}%
\affiliation{\utphysics}%

\author{Feliciano Giustino}
\email{fgiustino@oden.utexas.edu}
\affiliation{\oden}%
\affiliation{\utphysics}%

\maketitle
\onecolumngrid

\setlength{\parskip}{3pt}

\begin{tabular}{l@{\hskip 10pt}l}
\textbf{Contents} & \\[4pt]
Supplemental Figure~S1 & Transferability of symGW expressions across pseudopotentials and XC functionals\\[2pt]
Supplemental Figure~S2 & Band-gap renormalization of diamond compared to explicit GW calculations\\[2pt]
Supplemental Figure~S3 & Temperature dependence of the indirect gap of diamond\\[2pt]
Supplemental Figure~S4 & Comparison of symGW to explicit GW calculations for a-SiO$_2$\\[2pt]
\end{tabular}

\clearpage
\begin{figure}
    \centering
    \includegraphics[width=\linewidth]{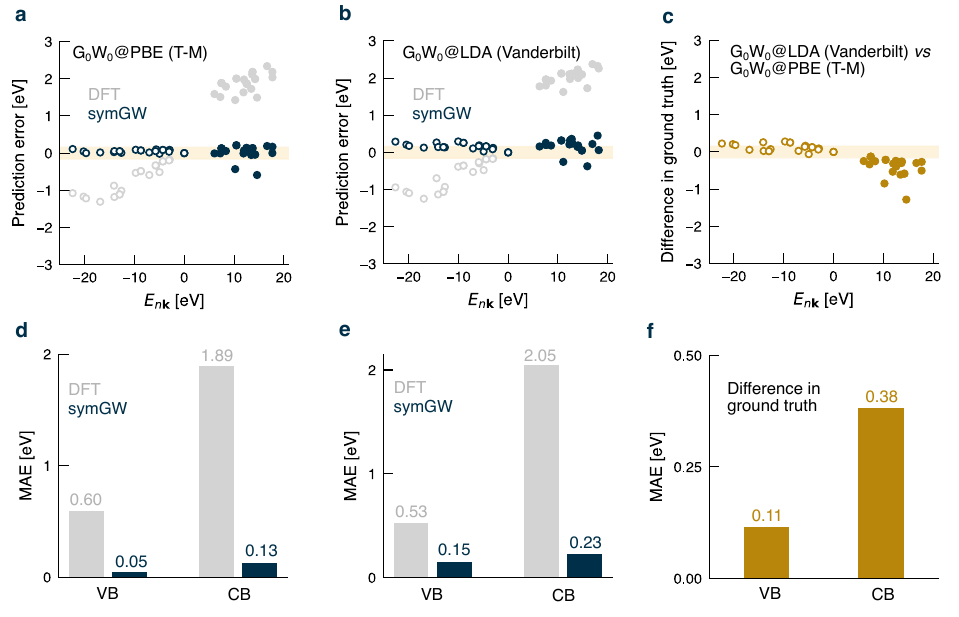}
    \caption{\textbf{Transferability of symGW expressions across pseudopotentials and XC functionals.} The transferability of symGW expressions derived from calculations using the PBE functional and a Vanderbilt norm-conserving pseudopotential [Eq. (1) of the main text] to \textbf{a}~calculations employing a different pseudopotential based on the Trouiller–Martins (T–M) scheme~\cite{PhysRevB.43.1993}, and 
    \textbf{b}~calculations using the local density approximation for the exchange–correlation functional. The prediction error is computed with respect to ground-truth GW calculations. 
    \textbf{c}~The difference in ground truth between the GW reference calculations of panels~a~and~b. For these comparisons, all energies are referenced to the valence-band top. 
    \textbf{d}~Mean absolute error (MAE) corresponding to the points in panel~a. 
    \textbf{e}~MAE corresponding to the points in panel~b. 
    \textbf{f}~MAE corresponding to the points in panel~c. Remarkably, the error in symGW predictions when applied to pseudopotentials and exchange–correlation functionals outside of the training set is comparable to or smaller than the discrepancy between the two independent ground-truth GW calculations.}
    \label{fig:transferrability}
\end{figure}

\clearpage
\begin{figure}
    \centering
    \includegraphics[width=0.8\textwidth]{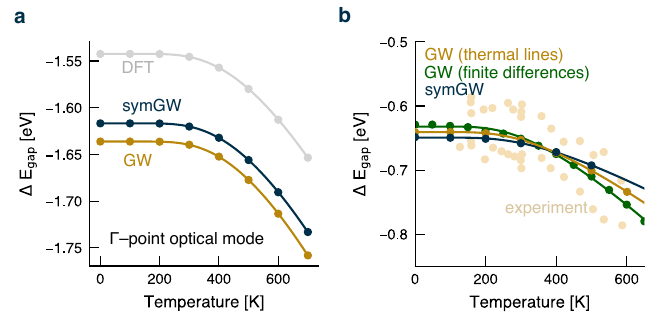}
    \caption{\textbf{Band-gap renormalization of diamond compared to explicit GW calculations.} \textbf{a} Comparison of DFT and symGW to explicit GW calculations for the renormalization of the direct band gap due to the $\Gamma$-point displacement pattern of an optical phonon mode in diamond. In this calculation, we consider a single $\textbf{k}$ point with unit weight, which effectively corresponds to replacing the entire optical branch with its $\Gamma$-point energy. symGW accurately reproduces both the shape and the absolute value of the band-gap renormalization. \textbf{b} Comparison of symGW band gaps computed in this work to published GW calculations along thermal lines~\cite{monserrat2016correlation} and using finite differences~\cite{antonius2014many} for the renormalization of the direct band gap. The zero-point renormalization with our method is $-$0.65 eV, which agrees with the previously reported values of $-$0.64~eV~\cite{monserrat2016correlation} and $-$0.63~eV~\cite{antonius2014many}. At 500~K, we find a band-gap renormalization of $-0.69$~eV, which also agrees with the previously reported values of $-0.70$~eV~\cite{monserrat2014extracting} and $-0.71$~eV~\cite{antonius2014many}. Experimental measurements are also shown for reference~\cite{logothetidis1992origin}. We do not consider the effects of lattice expansion with temperature.}
    \label{fig:placeholder}
\end{figure}

\clearpage
\begin{figure}
    \centering
    \includegraphics[width=0.5\textwidth]{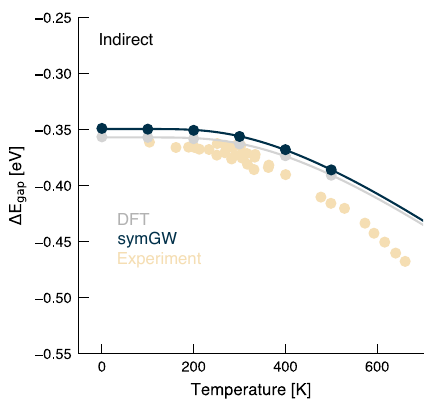}
    \caption{\textbf{Temperature dependence of the indirect bang gap of diamond.} Both DFT and symGW are in good agreement with experimental measurements of the indirect band gap~\cite{clark1964intrinsic}. The experimental quantum zero-point renormalization of $-0.36$~eV is obtained from isotope-shift measurements~\cite{cardona2005electron}. We do not consider the effects of lattice expansion with temperature. We calculate a zero-point renormalization of $-$0.36~eV at the DFT level and $-0.35$~eV at the GW level; these findings are consistent with previous calculations at the DFT level, which reported a value of $-0.33$ eV~\cite{ponce2015temperature}.}
    \label{fig:ind_gap_BGR}
\end{figure}

\clearpage
\begin{figure}
    \centering
    \includegraphics[width=0.9\linewidth]{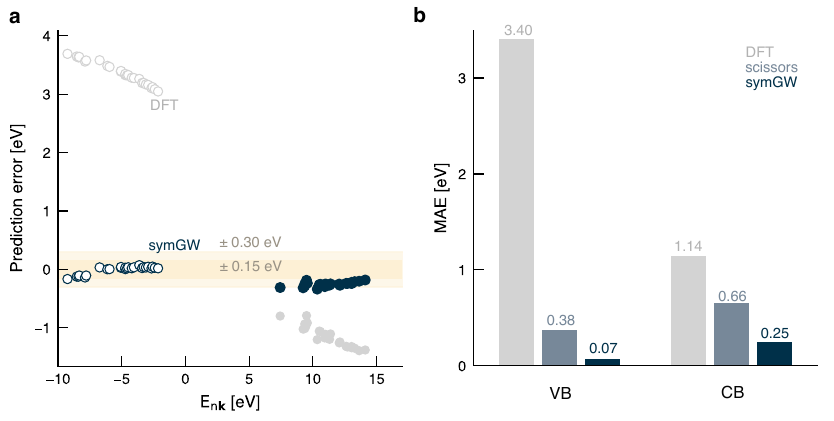}
    \caption{\textbf{Comparison of symGW to explicit GW calculations for a-SiO$_2$.} 
    \textbf{a} Prediction error of DFT and symGW for quasiparticle energies in a 72-atom a-SiO$_2$ supercell. Filled disks denote conduction states and open circles denote valence states.
    \textbf{b} Mean absolute error comparison between DFT, DFT with a rigid scissors shift, and symGW.} 
    \label{fig:aSiO2}
\end{figure}

\clearpage
\newpage

\bibliographystyle{apsrev4-2}
\bibliography{references}

%% file: PRB/appendix.tex
\colorlet{blue}{blue!70!black} 
\colorlet{red}{red!70!black} 
\newcommand{\cmvs}{$\text{cm}^2/\text{Vs}$}
\setlength{\parskip}{3pt}
\setlength{\parindent}{0pt}

\newpage
\section{Methods}\label{app:methods}
\smallskip

\textbf{Ground-state calculations.} We used \texttt{Quantum ESPRESSO}~\cite{giannozzi2017advanced, giannozzi2020quantum} to perform ground-state DFT calculations in the generalized-gradient approximation of Perdew, Burke, and Erznerhof~\cite{perdew1996generalized}, using optimized norm-conserving Vanderbilt pseudopotentials~\cite{hamann2013optimized} obtained from the \texttt{Pseudo Dojo} library~\cite{van2018pseudodojo}. For all materials, we relaxed the lattice constants and the internal atomic coordinates until all forces were below $10^{-3}$~Ry/bohr and the pressure was less than 0.5~kbar. To compute the charge density, we used a uniform Brillouin-zone (BZ) sampling grid of $12^3$ \textbf{k}-points for diamond, $12^2$ \textbf{k}-points for MoS$_2$, and $20^3$ \textbf{k}-points for $\alpha$-SiO$_2$. We used converged kinetic energy cutoffs of 100~Ry for diamond, 92~Ry for MoS$_2$, and 100~Ry for $\alpha$-SiO$_2$. We also tested a pseudopotential constructed with the Trouiller-Martins scheme for diamond, for which we used a kinetic energy cutoff of 250~Ry.

\textit{Electron--phonon renormalization in diamond}. To compute the electron--phonon renormalized band structure of diamond, we performed DFPT calculations on a BZ sampling grid of $8^3$ \textbf{q}-points. We used the Zacharias--Giustino (ZG) special displacement method~\cite{zacharias2020theory}, which generates a supercell with atoms distorted according to the thermal population of phonon modes. We calculated the direct and indirect band gaps in supercells by evaluating the energy separation between the peaks of the unfolded spectral function. 

\textit{Strained MoS$_2$}. For all calculations, we included spin–orbit coupling and applied a two-dimensional Coulomb truncation along the $z$ direction~\cite{sohier2017density}, using a vacuum-padded cell of 19.5~\AA\ along \textit{z}. To test the effects of strain, we uniformly sampled the three independent components of the 2D strain tensor ($\sigma_{11}, \sigma_{12}, \sigma_{22}$) using a Sobol sequence in the range $-3\%$ to $+3\%$, allowing the internal atomic coordinates to relax for each of the 24 strained configurations. In Fig.~3(c) of the main text, we report the harmonic mean of the conductivity effective mass~\cite{hautier2014does, ponce2020first}. 

\textit{Amorphous SiO$_2$}. We used a published 72-atom structural model of a-SiO$_2$ generated by a first-principles molecular dynamics quench from the melt~\cite{Sarnthein1995Model, giacomazzi2009medium, Giacomazzi2019VitreousSiO2}. This structural model is shown to closely reproduce the structure factor obtained from neutron diffraction experiments. To compute the charge density, we used a BZ sampling grid of $4^3$ \textbf{k}-points and a kinetic energy cutoff of 200~Ry. To compare the calculated electronic density of states to experiments in Fig. 3(f) of the main text, we broadened the delta function using a Gaussian spread of 1.5 eV corresponding to the resolution of the spectrometer~\cite{Fang1998Valence}.

\smallskip

\textbf{Many-body GW calculations.}
We used \texttt{BerkeleyGW}~\cite{deslippe2012berkeleygw} to apply many-body quasiparticle corrections within the $G_0W_0$ approximation and the generalized plasmon-pole approximation of Hybertsen and Louie~\cite{hybertsen1987ab}. We generated wave functions on uniform BZ sampling grids of $8^3$ \textbf{k}-points for diamond, $3^3$ \textbf{k}-points for the 54-atom ZG supercell of diamond, $12^2$ \textbf{k}-points for MoS$_2$, $4^3$ \textbf{k}-points for $\alpha$-SiO$_2$, and the $\Gamma$-point for a-SiO$_2$. For the dielectric function, we used a kinetic energy cutoff of 35~Ry for diamond, $\alpha$-SiO$_2$, and a-SiO$_2$, and 25~Ry for MoS$_2$.  For the summation over empty bands, we included states with energies up to 26~Ry for diamond (29~Ry for the calculation using the Trouiller-Martin pseudopotential), 18~Ry for the 54-atom ZG supercell of diamond, 5~Ry for MoS$_2$, 22~Ry for $\alpha$-SiO$_2$, and 9~Ry for a-SiO$_2$, all referenced to the respective valence-band top.

\smallskip

\textbf{Symbolic GW.}
To search for symbolic expressions, we used the evolutionary search implemented in the open source python library \texttt{PySR}~\cite{cranmer2023interpretable}. The training set comprises the KS4 vectors evaluated on the same BZ sampling grid used to compute the expectation values of the GW self energy. We find the best results when the conduction and valence manifolds are trained separately. For diamond [Eq. (1) of the main text] and $\alpha$-SiO$_2$ [Eq. (4) of the main text], we trained the models on quasiparticle corrections to the four highest valence bands and four lowest conduction bands. For MoS$_2$ [Eq. (3) of the main text], we trained the models on the two highest valence bands and two lowest conduction bands. 

\textit{Computational setup}. During the evolutionary search, we limited the search space to polynomials by restricting the set of binary operations to ($+,-,\times$) and the set of unary operations to the square and the cube. This choice is heuristically motivated by the fact that the quasiparticle self-energy admits a power-series expansion near quasiparticle peaks. We restricted the maximum complexity of the binary expression trees to 30, and found that increasing the complexity beyond this slows the search without improving the results. We set the complexity of constants to the default value of 1 and set the parsimony parameter to 0.01. We performed the search on 5 local processors and set up 10 distinct populations, each containing the default 27 individuals that evolve simultaneously. We ran the search for 100 iterations, with migration between populations at the end of each iteration. We used mini batching with batch sizes of 100 to accelerate the search process. All other parameters were set to their default values. We have found that this setup is robust and performs consistently across a wide range of semiconductors and insulators.

\clearpage
\onecolumngrid
\smallskip

\begin{table}[hp!]
\caption{\textbf{Candidate expressions for the conduction manifold of diamond.} All numbers are rounded to 3 decimal places. The expressions are reproduced here exactly as outputted by \texttt{pySR}, without any further algebraic simplification. All variables carry the $n\textbf{k}$ subscript, which we omit here for notational simplicity. Hartree atomic units are used. The error is computed for all states in the four lowest conduction bands of diamond in a BZ grid with $8\times8\times8$ points.}
\vspace{10pt}
\centering
\begin{tabular}{@{\extracolsep{\fill}} l @{\hspace{0.5cm}} l @{\hspace{0.5cm}} r}
\hline\\[-9.5pt] \hline\\[-5pt]
Complexity & Expression & MAE\\
 & & eV \\[2pt]
 &  & \\[-8pt]
\hline\\[-5pt]
1 & $v_\text{xc}$ & 0.536\\
3 & $v_\text{xc}\times 0.964$ & 0.152\\
5 & $v_\text{xc}\times 0.805 - 1\times 0.0914$ & 0.088\\
6 & $v_\text{xc} - (v_\text{xc} + 0.302)^3$ & 0.085\\
7 & $v_\text{H} + v_\text{xc}\times 3.461 + 1.197$ & 0.061\\
9 & $v_\text{H}\times 0.885 + v_\text{xc}\times 3.156 + 1.049$ & 0.062\\
12 & $-\varepsilon^{12} + v_\text{H} + v_\text{xc}\times 3.456 + 1.194$ & 0.061\\
13 & $-(-0.013)\times \varepsilon + v_\text{H}\times 0.565 + v_\text{xc}\times 2.333 + 0.647$ & 0.061\\
15 & $v_\text{H}\times 0.859 + (v_\text{xc} - (0.515 - \varepsilon)^6)\times 3.102 + 1.024$ & 0.057\\
17 & $v_\text{H}\times 0.859 + v_\text{xc}\times 3.102 - 1.872\times (-0.901\times \varepsilon + 0.901\times v_\text{xc} + 1)^6 + 1.024$ & 0.055\\
20 & $-(\varepsilon + v_\text{xc})\times (\varepsilon + v_\text{H} + v_\text{xc})^3 + (v_\text{H}\times 0.271 + v_\text{xc})\times 2.968 + 0.960$ & 0.056\\
22 & $v_\text{H}\times 0.820 + v_\text{xc}\times 3.013 - (\varepsilon + v_\text{xc})\times (\varepsilon + v_\text{H} + v_\text{xc} - 1\times 0.008)^3 + 0.982$ & 0.055\\
23 & $v_\text{H}\times 0.852 + v_\text{xc}\times 3.099 - (\varepsilon + v_\text{xc})^3\times ((\varepsilon + v_\text{H} + v_\text{H})^3 - 0.580)^3 + 1.023$ & 0.054\\
25 & $v_\text{H}\times 0.682 + v_\text{xc}\times 2.668 - ((\varepsilon + v_\text{H} + v_\text{H})^3 - 0.569)^3\times (\varepsilon + v_\text{xc} - 1\times 0.038)^3 + 0.817$ & 0.056\\[3pt]
\hline\\[-9.5pt] \hline\\[-7pt]
\label{tab:diamond_conduction}
\end{tabular}
\end{table}

\clearpage

\begin{table}
\caption{\textbf{Candidate expressions for the valence manifold of diamond.} All numbers are rounded to 3 decimal places. The expressions are reproduced here exactly as outputted by \texttt{pySR}, without any further algebraic simplification. All variables carry the $n\textbf{k}$ subscript, which we omit here for notational simplicity. Hartree atomic units are used. The error is computed for all states in the four valence bands of diamond in a BZ grid with $8\times8\times8$ points.}
\vspace{10pt}
\centering
\begin{tabular}{@{\extracolsep{\fill}} l @{\hspace{0.5cm}} l @{\hspace{0.5cm}} r}
\hline\\[-9.5pt] \hline\\[-5pt]
Complexity & Expression & MAE\\
 & & eV \\[2pt]
 &  & \\[-8pt]
\hline\\[-5pt]
1 & $-0.655$ & 0.286\\
4 & $-v_\text{H}^3 - 0.640$ & 0.255\\
5 & $v_\text{xc}\times 0.454 - 0.392$ & 0.232\\
6 & $v_\text{H}^3 + v_\text{xc} - 1\times 0.091$ & 0.232\\
7 & $v_\text{xc}\times (1.084 - 0.141\times \varepsilon)$ & 0.063\\
9 & $-(-0.420)\times \varepsilon\times v_\text{H} + v_\text{xc} - 0.045$ & 0.041\\
11 & $-(-0.325)\times \varepsilon\times (v_\text{H} + 0.053) + v_\text{xc} - 0.046$ & 0.037 \\
12 & $-\varepsilon^2\times v_\text{H}\times (\varepsilon + 1.170) + v_\text{xc} - 0.055$ & 0.031\\
13 & $v_\text{xc}^4\times (\varepsilon^6 + \varepsilon - 1.369) - 1\times 0.466$ & 0.024\\
14 & $v_\text{xc}^4\times (\varepsilon + (-\varepsilon - 0.187)^3 - 1.338) - 0.469$ & 0.017\\
16 & $v_\text{xc}^4\times (\varepsilon + (\varepsilon^2 + v_\text{H}^2)^3 - 1.344) - 0.469$ & 0.017\\
18 & $v_\text{xc}^4\times (\varepsilon + (\varepsilon - 4.081\times v_\text{H}^3)^6 - 1.346) - 1\times 0.469$ & 0.016 \\
19 & $v_\text{xc}^4\times (\varepsilon + (\varepsilon - (v_\text{H} + v_\text{H})^4)^6 - 1.352) - 1\times 0.469$ & 0.016\\
20 & $v_\text{xc}^4\times (\varepsilon + (\varepsilon - (v_\text{H} + v_\text{H} - 1\times 0.220)^2)^6 - 1.350) - 0.469$ & 0.015\\[3pt]
\hline\\[-9.5pt] \hline\\[-7pt]
\label{tab:diamond_valence}
\end{tabular}
\end{table}

\clearpage
\begin{table}
\caption{\textbf{Self-energy as rescaled XC energy}. We quantify the error made when approximating the self-energy as a rescaled XC energy, $\Sigma_{n\textbf{k}} = \alpha_1 v^\text{XC}_{n\textbf{k}} + \alpha_2$. Values for these parameters, obtained from a least-squares regression, and the corresponding mean absolute errors and maximum absolute errors in the Brillouin zone are shown. These errors are evaluated on uniform k-grids over the indicated energy ranges. Mean absolute errors below 0.1~eV are shown in bold.}
\vspace{10pt}
\centering
\begin{tabular}{@{\extracolsep{\fill}} l @{\hspace{0.5cm}} l @{\hspace{0.5cm}} r @{\hspace{0.5cm}} r @{\hspace{0.5cm}} r @{\hspace{0.5cm}} r @{\hspace{0.5cm}} r}
\hline\\[-9.5pt] \hline\\[-5pt]
Material & Manifold & $\alpha_1$ & $\alpha_2$ & Mean Error & Max Error & Energy Range\\
 &  &  & Ha & eV & eV & eV\\[2pt]
 & & & & & & \\[-8pt]
\hline\\[-5pt]
Diamond & Conduction & 0.805 & $-$0.0912 & \textbf{0.088} & 0.459 & $[\varepsilon_\text{CBM},\  \varepsilon_\text{CBM}+12.5]$\\
Diamond & Valence & 0.466	& $-$0.385 & 0.234 & 0.647 & $[\varepsilon_\text{VBM} - 21.4, \ \varepsilon_\text{VBM}]$ \\
MoS$_2$ (2D) & Conduction & 1.000 & 0.0317 & 0.268 & 0.744 & $[\varepsilon_\text{CBM},\  \varepsilon_\text{CBM}+9.49]$ \\
MoS$_2$ (2D) & Valence & 1.000 & $-$0.0131 & 0.265 & 0.706 & $[\varepsilon_\text{VBM} - 34.2, \ \varepsilon_\text{VBM}]$ \\
$\alpha$-SiO$_2$ & Conduction & 0.966 & 0.0217 & 0.101 & 0.443 & $[\varepsilon_\text{CBM},\  \varepsilon_\text{CBM}+9.84]$ \\
$\alpha$-SiO$_2$ & Valence & 0.844 & $-$0.195 & 0.288 & 0.917 & $[\varepsilon_\text{VBM} - 17.15, \ \varepsilon_\text{VBM}]$\\
Si  & Conduction & 0.958 & 0.195 & \textbf{0.027} & 0.100 & $[\varepsilon_\text{CBM},\  \varepsilon_\text{CBM}+10.4]$\\
Si  & Valence & 0.996 & 0.182 & \textbf{0.042} & 0.252 & $[\varepsilon_\text{VBM} - 11.5, \ \varepsilon_\text{VBM}]$\\
3C-SiC & Conduction & 0.864 & 0.285 & \textbf{0.051} & 0.370 & $[\varepsilon_\text{CBM},\  \varepsilon_\text{CBM}+14.9]$\\
3C-SiC & Valence & 0.914 & 0.241 & 0.129 & 0.375 & $[\varepsilon_\text{VBM} - 15.7, \ \varepsilon_\text{VBM}]$\\
GaAs & Conduction & 1.058 & 0.338 & \textbf{0.067} & 0.140 & $[\varepsilon_\text{CBM},\ \varepsilon_\text{CBM} + 11.0]$\\
GaAs & Valence & 0.875 & 0.210 & 0.119 & 0.437 & $[\varepsilon_\text{VBM} - 13.2,\ \varepsilon_\text{VBM}]$\\
WSe$_2$ (2D) & Conduction & 0.931 & $-$0.188 & 0.118 & 0.534 & $[\varepsilon_\text{CBM},\ \varepsilon_\text{CBM} + 8.04]$\\
WSe$_2$ (2D)  & Valence & 0.596 & $-$0.431 & 0.156 & 0.556 & $[\varepsilon_\text{VBM} - 10.1,\ \varepsilon_\text{VBM}]$\\
WTe$_2$ (2D) & Conduction & 0.806 & $-$0.240 & 0.109 & 0.501 & $[\varepsilon_\text{CBM},\ \varepsilon_\text{CBM} + 2.16]$\\
WTe$_2$ (2D)  & Valence & 0.655 & $-$0.360 & 0.101 & 0.336 & $[\varepsilon_\text{VBM} - 9.00,\ \varepsilon_\text{VBM}]$ \\
LiF  & Conduction & 0.950 & 0.075 & 0.250 & 0.632 & $[\varepsilon_\text{CBM},\ \varepsilon_\text{CBM} + 21.7]$\\
LiF  & Valence & 1.301 & 0.095 & 0.771 & 1.431 & $[\varepsilon_\text{VBM} - 22.3,\ \varepsilon_\text{VBM}]$\\
MgO  & Conduction & 0.966 & 0.303 & 0.246 & 0.629 & $[\varepsilon_\text{CBM},\ \varepsilon_\text{CBM} + 18.1]$\\
MgO  & Valence & 1.177 & 0.268 & 0.599 & 1.249 & $[\varepsilon_\text{VBM} - 18.9,\ \varepsilon_\text{VBM}]$\\[3pt]
\hline\\[-9.5pt] \hline\\[-7pt]
\label{tab:rescaled_XC}
\end{tabular}
\end{table}

\clearpage
\begin{table}
\caption{\textbf{Self-energy as rescaled XC energy plus dynamical corrections}. The error in modeling the self-energy as a rescaled XC energy plus dynamical corrections, $\Sigma_{n\textbf{k}} = \alpha_1 v^\text{XC}_{n\textbf{k}} + \alpha_2 + \alpha_3 \varepsilon_{n\textbf{k}} + \alpha_4 t_{n\textbf{k}}$. Values for these parameters, obtained from a least-squares regression, and the corresponding mean absolute errors and maximum absolute errors in the Brillouin zone are shown. These errors are evaluated on uniform k-grids over the indicated energy ranges. Mean absolute errors below 0.1~eV are shown in bold.}
\vspace{10pt}
\centering
\begin{tabular}{@{\extracolsep{\fill}} l @{\hspace{0.5cm}} l @{\hspace{0.5cm}} r @{\hspace{0.5cm}} r @{\hspace{0.5cm}} r @{\hspace{0.5cm}} r @{\hspace{0.5cm}} r @{\hspace{0.5cm}} r @{\hspace{0.5cm}} r}
\hline\\[-9.5pt] \hline\\[-5pt]
Material & Manifold & $\alpha_1$ & $\alpha_2$ &  $\alpha_3$  &  $\alpha_4$ & Mean Error & Max Error & Energy Range\\
 &  &  & Ha &  &  & eV & eV & eV\\[2pt]
 & & & & & & & & \\[-8pt]
\hline\\[-5pt]
Diamond & Conduction & 0.871 & $-$0.0655 & 0.0308 & 0.000 & \textbf{0.063} & 0.470 & $[\varepsilon_\text{CBM},\  \varepsilon_\text{CBM}+12.5]$ \\
Diamond & Valence & 1.163	& 0.0454 & 0.0833 & 0.000 &  \textbf{0.070} & 0.523 &  $[\varepsilon_\text{VBM} - 21.4, \ \varepsilon_\text{VBM}]$\\

MoS$_2$ (2D) & Conduction & 1.000 & $-$0.0125 & $-$0.003 & 0.026 &  \textbf{0.046} & 0.226 & $[\varepsilon_\text{CBM},\  \varepsilon_\text{CBM}+9.49]$ \\
MoS$_2$ (2D) & Valence & 1.000 & $-$0.0544 & 0.006 & 0.0294 &  \textbf{0.053} & 0.171 & $[\varepsilon_\text{VBM} - 34.2, \ \varepsilon_\text{VBM}]$\\

$\alpha$-SiO$_2$ & Conduction & 0.901 & $-$0.0386 & 0.113 & $-$0.002 &  \textbf{0.043} & 0.151 & $[\varepsilon_\text{CBM},\  \varepsilon_\text{CBM}+9.84]$ \\
$\alpha$-SiO$_2$ & Valence & 0.976 & $-$0.129 & 0.102 & 0.000 &  \textbf{0.020} & 0.107 & $[\varepsilon_\text{VBM} - 17.1, \ \varepsilon_\text{VBM}]$\\

Si  & Conduction & 0.970 & 0.198 & 0.005 & 0.000 & \textbf{0.025} & 0.112 & $[\varepsilon_\text{CBM},\  \varepsilon_\text{CBM}+10.4]$\\
Si  & Valence & 1.005 & 0.182 & $-$0.013 & 0.000 &  \textbf{0.031} & 0.171 & $[\varepsilon_\text{VBM} - 11.5, \ \varepsilon_\text{VBM}]$\\

3C-SiC & Conduction & 0.957 & 0.317 & 0.032 & 0.000 & \textbf{0.030} & 0.190 & $[\varepsilon_\text{CBM},\  \varepsilon_\text{CBM}+14.9]$\\
3C-SiC & Valence & 0.996 & 0.288 & 0.037 & 0.000 & \textbf{0.038} & 0.245 & $[\varepsilon_\text{VBM} - 15.7, \ \varepsilon_\text{VBM}]$\\

GaAs & Conduction & 1.064 & 0.336 & 0.027 & 0.000 & \textbf{0.040} & 0.175 & $[\varepsilon_\text{CBM},\ \varepsilon_\text{CBM} + 11.0]$\\
GaAs & Valence & 0.841 & 0.204 & 0.057 & 0.000 & \textbf{0.075} & 0.271 & $[\varepsilon_\text{VBM} - 13.2,\ \varepsilon_\text{VBM}]$\\

WSe$_2$ (2D) & Conduction & 1.050 & $-$0.188 & 0.010 & 0.043 & \textbf{0.076} & 0.269 & $[\varepsilon_\text{CBM},\ \varepsilon_\text{CBM} + 8.04]$\\
WSe$_2$ (2D)  & Valence & 1.272 & $-$0.169 & $-$0.038 & 0.092 & \textbf{0.047} & 0.188 & $[\varepsilon_\text{VBM} - 10.1,\ \varepsilon_\text{VBM}]$\\

WTe$_2$ (2D) & Conduction & 1.166 & $-$0.141 & 0.011 & 0.056 &  \textbf{0.038} & 0.156 & $[\varepsilon_\text{CBM},\ \varepsilon_\text{CBM} + 2.16]$\\
WTe$_2$ (2D)  & Valence & 1.143 & $-$0.186 & $-$0.017 & 0.061 &  \textbf{0.050} & 0.191 & $[\varepsilon_\text{VBM} - 9.00,\ \varepsilon_\text{VBM}]$ \\

LiF  & Conduction & 0.873 & $-$0.010 & 0.086 & 0.000 &  \textbf{0.076} & 0.313 & $[\varepsilon_\text{CBM},\ \varepsilon_\text{CBM} + 21.7]$\\
LiF  & Valence & 0.963 & $-$0.150 & 0.012 & 0.000 & \textbf{0.014} & 0.043 & $[\varepsilon_\text{VBM} - 22.3,\ \varepsilon_\text{VBM}]$\\

MgO  & Conduction & 0.942 & 0.251 & 0.079 & 0.000 & \textbf{0.084} & 0.187 & $[\varepsilon_\text{CBM},\ \varepsilon_\text{CBM} + 18.1]$\\
MgO  & Valence & 1.047 & 0.203 & 0.115 & 0.000 & \textbf{0.020} & 0.079 & $[\varepsilon_\text{VBM} - 18.9,\ \varepsilon_\text{VBM}]$\\[3pt]
\hline\\[-9.5pt] \hline\\[-7pt]
\label{tab:linear_selfen}
\end{tabular}
\end{table}

\clearpage
\newpage